# Sleep effects on brain, cognition, and mental health during adolescence are mediated by the glymphatic system

Xinglin Zeng[1], Yiran Li[1], Fan Nils Yang[2], Gianpaolo Del Mauro[1], Jiaao Yu[3], Ruoxi Lu[1], Jiachen Zhuo[1], Laura Rowland[4], Wickwire Emerson[5], Ze Wang[1,4]

1. Department of Diagnostic Radiology and Nuclear Medicine, University of Maryland School of Medicine,
2. NINDS, National Institutes of Health,
3. Department of Mathematics, University of Maryland College Park,
4. Maryland Psychiatry Research Center, Department of Psychiatry, University of Maryland School of Medicine,
5. Department of Psychiatry, University of Maryland School of Medicine,

Correspondence: ze.wang@som.umaryland.edu

Keywords: sleep, the glymphatic system, adolescence, brain structure, cognition, pathophysiology, behavior

Abstract

**Background:** Adolescence is a critical period of brain maturation and heightened vulnerability to cognitive and mental health disorders. Sleep plays a vital role in neurodevelopment, yet the mechanisms linking insufficient sleep to adverse brain and behavioral outcomes remain unclear. The glymphatic system (GS), a brain-wide clearance pathway, may provide a key mechanistic link.

**Methods:** Leveraging baseline data from the Adolescent Brain Cognitive Development (ABCD) Study, we examined whether GS function mediates the effects of sleep on brain structure, cognition, and mental health. GS function was indexed by perivascular space (PVS) burden derived from structural MRI. Participants (n ≈ 6,800; age ≈ 11 years) were categorized into sleep-sufficient (≥9 h/night) and sleep-insufficient (<9 h/night) groups. Linear models tested associations among sleep, PVS burden, brain volumes, and behavioral outcomes. Mediation analyses evaluated whether PVS burden explained sleep-related effects.

**Results:** Adolescents with insufficient sleep exhibited significantly greater PVS burden (Cohen's d ≈ 0.15), reduced cortical, subcortical, and white matter volumes, poorer cognitive performance across multiple domains (largest effect in crystallized intelligence, d ≈ 0.20), and elevated psychopathology (largest effect in general problems, d ≈ –0.34). Sleep duration and quality were strongly associated with PVS burden ($p < 10^{-8}$). Mediation analyses revealed that PVS burden partially mediated sleep effects on cognition (e.g., crystallized intelligence, episodic memory) and mental health (e.g., psychosis severity), with indirect proportions up to 10.9%. Sequential models suggested a pathway from sleep → PVS → brain volume → behavior as the most plausible route.

**Conclusions:** Insufficient sleep during adolescence is linked to glymphatic dysfunction, reflected by increased PVS burden, which partially accounts for adverse effects on brain structure, cognition, and mental health. These findings highlight the glymphatic system as a potential mechanistic pathway and imaging biomarker, underscoring the importance of promoting adequate sleep to support neurodevelopment and mental health.

**Introduction**

Adolescence is a period of rapid brain growth and cognitive development, accompanied by heightened vulnerability to mental disorders(Bale et al., 2010; Curtis, 1992; DiClemente et al., 2013; Dumontheil et al., 2008; Giedd, 2004; Paus et al., 2008; Schmaal et al., 2017; Turner et al., 1993). Identifying health risk factors and their underlying neurobiological mechanisms during this sensitive developmental window is essential for optimizing mental health and designing early interventions. Sleep has emerged as a critical public health concern among adolescents, with a rising prevalence of insufficient and fragmented sleep(Twenge et al., 2017). National survey data from 2015 indicate that 40% of US adolescents obtained fewer than seven hours of sleep per night(Twenge et al., 2017), a pattern that may have enduring consequences given the accelerated neurodevelopmental processes occurring during this period. Sleep serves essential homeostatic and regulatory functions(Hobson, 2005), including protein synthesis, hormonal modulation, and clearance of metabolic waste(Czeisler & Klerman, 1999; Imeri & Opp, 2009; Mignot, 2008; Robles & Carroll, 2011; Xie et al., 2013). It also facilitates neural reorganization critical for high-order cognitive functions such as executive control, learning, and memory consolidation(Abel et al., 2013; Diekelmann & Born, 2010; Karni et al., 1994; Stickgold, 2005). As the body, the brain, and cognitive function undergo continued maturation throughout adolescence, this developmental stage may be particularly susceptive to the deleterious effects of inadequate or poor-quality sleep(Gregory & Sadeh, 2016; Telzer et al., 2013). Repeated disruptions in sleep—whether through reduced duration or impaired quality--may lead to long-lasting or irreversible alterations in brain structure and function, with downstream consequences for cognitive, behavioral, and mental health outcomes(Beebe, 2011; Medic et al., 2017; Owens et al., 2014; Tarokh et al., 2016; Turnbull et al., 2013).

Consistent with these observations, our recent studies have identified widespread negative effects of insufficient sleep on brain morphology, connectivity, cognition, and mental health among adolescents(Turan et al., 2025; Yang et al., 2021; Yang, Liu, et al., 2022; Yang et al., 2023; Yang, Xie, et al., 2022a, 2022b). These effects persisted two years(Yang, Xie, et al., 2022a), raising concerns about their cumulative and irreversible nature(Beebe, 2011; Medic et al., 2017; Owens et al., 2014; Tarokh et al., 2016; Turnbull et al., 2013). These findings have been reported by NIH and >170 international media outlets, underscoring the broad public and scientific importance of adolescent sleep research.

Despite this scientific importance and public health interest, the neurobiological mechanisms underlying these profound effects of sleep remain poorly understood. Based on several lines of converging evidence, we propose to investigate the GS as a compelling mechanistic pathway linking sleep to brain, cognitive, and psychopathological (BCP) outcomes. First, the brain's exceptionally high energy consumption results in the continuous production of metabolic waste, including neurotoxic proteins such as amyloid-beta and tau proteins. Efficient clearance of these metabolic byproducts is essential for maintaining brain homeostasis and function. Notably, the adolescent brain has even higher energy metabolism than the adult brain, resulting in an increased demand for effective clearance of metabolic waste products. Coinciding with this metabolic surge, humans exhibit the longest sleep duration prior to adulthood, further suggesting an elevated physiological need for sleep-mediated clearance processes during this developmental period. Second, the GS(Iliff et al., 2014; Iliff et al., 2013; Iliff & Nedergaard, 2013; Iliff et al., 2012; Kress et al., 2014; Nedergaard, 2013)—a recently discovered, glial cell-modulated, brain-wide lymphatic-like clearance system—has been identified as a major pathway for the removal of brain waste products(Iliff et al., 2014; Iliff et al., 2013; Iliff & Nedergaard, 2013; Iliff et al., 2012; Kress et al., 2014; Nedergaard, 2013). The GS facilitates cerebrospinal fluid (CSF) and interstitial fluid (ISF) exchange through perivascular pathways, promoting the clearance of metabolic waste from the brain parenchyma. Its discovery represents a significant breakthrough in neuroscience, resolving a longstanding mystery regarding how the brain clears waste(Benveniste et al., 2019; Bohr et al., 2022; Harrison et al., 2020; Iliff et al., 2013; Iliff & Nedergaard, 2013; Iliff et al., 2012; Kamagata et al., 2022; Nedergaard & Goldman, 2020; Peng et al., 2016; Rasmussen et al., 2018; Taoka et al., 2017). Third, accumulating evidence from human and animal studies indicates that GS activity peaks during sleep and is causally impaired by sleep disruption(Achariyar et al., 2016; Eide et al., 2021; Fultz et al., 2019; Hauglund et al., 2025; Hauglund et al., 2020; Lee et al., 2015; Levendowski et al., 2019; Mistlberger & Kent, 2019; Miyakoshi et al., 2023; Reddy & van der Werf, 2020; Shokri-Kojori et al., 2018; Siow et al., 2022; Vasciaveo et al., 2023; Xie et al., 2013; Zhang et al., 2020). Moreover, GS dysfunction has been found to lead to or be related to toxic protein accumulations(Boespflug & Iliff, 2018; Harrison et al., 2020; Iliff et al., 2014; Kress et al., 2014; Nedergaard & Goldman, 2020; Tarasoff-Conway et al., 2015; Wang et al., 2021; Zeppenfeld et al., 2017), vascular pathology(Mestre et al., 2017; Tang et al., 2022), cognitive decline(Kamagata et al., 2022; Paradise et al., 2021; Reeves et al., 2020; Wang et al., 2022), autism(Li et al., 2022), depression(Chen et al., 2021; Liu et al., 2020; Ranti et al., 2022; Xia et al., 2017), alcohol use disorder(Lundgaard et al., 2018), etc.

Although GS has not been systematically examined during adolescence, the converging evidence from human adults and animal models clearly underscores the importance of investigating the relationship between GS and sleep in this critical developmental stage. Furthermore, while prior research has primarily focused on the GS's role in waste clearance, emerging studies suggest that it also contributes to other fundamental neurobiological processes vital for brain development and function. These include water homeostasis (Hussain et al., 2023; Mestre et al., 2020; Rasmussen et al., 2022), antigen presentation(Louveau et al., 2017), immune cell trafficking(Kipnis, 2024; Louveau et al., 2018), and in the delivery of nutrients, such as glucose(Lundgaard et al., 2015). Although the sensitivity of these additional GS functions to sleep remains unknown, their importance during adolescence further underscores the need for comprehensive investigation of GS activity and its relationship with sleep during this vulnerable period.

This study represents a first-of-its-kind effort to address this critical gap using the publicly available large-scale sleep, imaging, and outcomes data from the Adolescent

Brain and Cognitive Development (ABCD) project. We used structural MRI-derived perivascular space (PVS) volume as the proxy measure of the GS. We first examined the effects of sleep duration on PVS and outcome measures and then used mediation analysis to illustrate the modulating effects of PVS on the sleep effects on neurocognitive and behavioral outcome. Our hypotheses are: shorter sleep duration is associated with larger PVS volume; sleep duration effects on cognition and behavior are significantly mediated by PVS.

## Method

### Study design and data source

This study leveraged data from the Adolescent Brain Cognitive Development (ABCD) Study, a nationwide, longitudinal research initiative designed to investigate brain development and health trajectories in U.S. youth (Casey et al., 2018). The ABCD Study was conducted in accordance with the ethical principles outlined in the Declaration of Helsinki. Written informed consent was obtained from a parent or legal guardian, and assent was provided by each child participant. The protocol was approved by the Institutional Review Boards (IRBs) at each of the 21 participating research sites.

Participants were included in the current analysis if they met the following criteria: (1) availability of high-quality T1-weighted and T2-weighted structural MRI scans; (2) complete sleep data; and (3) available demographic information, including age and sex. Sleep metrics were derived from the Modified Munich ChronoType Questionnaire (MCTQ), which assesses both weekday and weekend sleep behaviors (Rohr et al., 2025). Habitual sleep duration was estimated by calculating a weighted average of weekday and weekend sleep hours. Then participants were categorized into sleep-sufficient (SS; n = 4,257, age: 11.435 ± 0.501) and sleep-insufficient (SI; n = 2,571, age 11.519 ± 0.507; ) groups based on a 9-hour cutoff consistent with recommended sleep guidelines for adolescents (Yang, Xie, et al., 2022a). In addition, subjective sleep quality was assessed using the parent-reported Sleep Disturbance Scale for Children (SDSC). The demographic data was presented in Table S1.

**MRI data acquisition**

he multi-site structural MRI data were collected using a harmonized imaging protocol across 21 sites, employing 3T MRI scanners from three vendors: Siemens Prisma, GE 750, and Philips (Casey et al., 2018). All scans were performed using standard adult-size multi-channel head coils and adhered to a unified acquisition protocol to ensure inter-site compatibility. High-resolution T1-weighted images were acquired using a 3D magnetization-prepared rapid gradient-echo (MPRAGE) sequence with the following parameters: repetition time (TR) = 2500 ms, echo time (TE) = 2.88 ms, inversion time (TI) = 1060 ms, flip angle = 8°, field of view (FOV) = 256 mm × 256 mm, matrix size = 256 × 256, and slice thickness = 1.0 mm, yielding isotropic voxels of 1.0 mm³. T2-weighted images were collected using a 3D variable flip angle turbo spin-echo sequence with the following parameters: TR = 3200 ms, TE = 565 ms, flip angle = variable, FOV = 256 mm × 256 mm, matrix size = 256 × 256, and slice thickness = 1.0 mm, also providing 1.0 mm³ isotropic resolution. All acquired images underwent prospective motion correction to minimize motion artifacts, followed by quality control and centralized preprocessing.

**MRI data preprocess**

T1-weighted and T2-weighted structural MRI images were preprocessed using the Human Connectome Project (HCP) minimal preprocessing pipeline (Glasser et al., 2013). The preprocessing steps included gradient nonlinearity correction, readout distortion correction, bias field correction, and alignment to the anterior commissure–posterior commissure (AC-PC) axis. Skull stripping and tissue segmentation were performed using FreeSurfer, and individual structural images were registered to the MNI152 standard space using non-linear registration. To enhance PVS visibility, adaptive non-local mean filtering was applied to both T1-weighted and T2-weighted images to reduce high-frequency noise while preserving fine structural details. The filtered T1w and T2w images were then used to compute the Enhanced PVS Contrast (EPC) map by voxel-wise division of the filtered T1w image by the T2w image (Farshid Sepehrband et al., 2019; Jiachen Zhuo et al., 2024).

Brain volumetric data were extracted from FreeSurfer's aseg.stats files, generated by the recon-all pipeline applied to preprocessed T1-weighted images. We extracted total volumes of cortical gray matter, cerebral white matter, and subcortical gray matter to serve as measures of brain tissue compartments for subsequent mediation analyses.

**Perivascular space segmentation**

PVS segmentation was performed using a vessel-likeliness filtering approach. Specifically, T1-weighted and T2-weighted images were processed using the Frangi vesselness filter, which identifies tubular structures by evaluating the eigenvalues of the Hessian matrix. White matter masks were derived using the ANTs n-tissue parcellation framework, and PVS segmentation was restricted to these white matter regions. Vesselness maps were thresholded to generate binary PVS masks (F. Sepehrband et al., 2019). The threshold was empirically optimized to match expert visual ratings, with a final vesselness threshold of 0.0000001 used for segmentation. To avoid potential false positives in regions not typically associated with PVS, the corpus callosum was excluded from the white matter mask. Normalized PVS volume (i.e., total PVS volume divided by total white matter volume), were extracted for statistical analysis (J. Zhuo et al., 2024).

In addition to individual-level burden estimation, we constructed a probabilistic PVS distribution map across the entire ABCD baseline cohort, including participants without sleep data (Figure S1). For each voxel in standard space, we computed the proportion of participants who exhibited segmented PVS at that location. This resulted in a voxel-wise probability map reflecting the spatial consistency of PVS across the population. Higher voxel values indicate greater inter-individual overlap and a higher likelihood of representing anatomically consistent PVS in adolescence.

To reduce the influence of spurious or participant-specific artifacts, voxels present in fewer than 0.05% of participants were excluded.

These probabilistic PVS maps serve dual purposes: (1) as anatomical priors to constrain voxel-wise analyses to regions with high reliability across individuals, and (2) as population-based templates for future research on glymphatic structure and function during adolescent brain development. The use of a large cohort enhances the generalizability and anatomical precision of these templates.

**Neurocognition, and mental health measurement**

Neurocognitive and mental health outcomes were assessed using standardized instruments from the Adolescent Brain Cognitive Development (ABCD) Study baseline dataset (Barch et al., 2021; Luciana et al., 2018). Cognitive functions assessments included: 1) Executive Function using the Go/No-Go task net score, which reflects response inhibition and attentional control, and the List Sorting Working Memory Test, which measures the capacity to maintain and manipulate information in working memory; 2) Crystallized Intelligence using the NIH Toolbox Crystallized Composite Score, which integrates performance on vocabulary and reading recognition tasks. 3) Episodic Memory using the Rey Auditory Verbal Learning Test (RAVLT), including both immediate recall and delayed recall subtests. 4) Social Cognition using the Social Interaction Task, designed to measure the ability to recognize and interpret social cues and emotions. 5) Visual Processing using the Snellen Visual Acuity test, with scores reflecting the clarity of visual perception. Higher values across all cognitive measures indicated better cognitive performance.

Mental health domains were assessed using validated parent- and youth-reported questionnaires: 1) General Psychopathology was measured using the Achenbach Adult Behavior Checklist (ABCL) Total Problems score and the Brief Problem Monitor – Youth (BPM-Y) total problems mean score. 2) Mood and Bipolar Symptoms were assessed using the Parent General Behavior Inventory – 10 Item Mania Scale (PGBI-10M). 3) Psychotic-like Symptoms were evaluated using the Prodromal Questionnaire – Brief Child Version (PQ-BC) total number of symptoms and the Psychosis Symptom Severity (PSS) score. 4) Impulsivity and Inhibition were measured using the Behavioral Inhibition System (BIS) sum score and the UPPS-P Negative Urgency subscale. 5) Reward Sensitivity was assessed via the Behavioral Activation System – Reward Responsiveness (BAS-RR) subscale. 6) Peer Problems were evaluated using the Peer Experiences Questionnaire (PEQ) Overall Mean score. Note that PSS is positively oriented (higher scores indicate better functioning), whereas all other mental health scales are symptom-oriented (higher scores indicate greater problems).

**Mediation analysis**

To examine whether PVS burden mediated the relationship between sleep characteristics and behavioral outcomes, we performed causal mediation analyses using the lavaan package in R (Rosseel, 2012). All models were estimated using structural equation modeling with standardized path coefficients and bias-corrected bootstrap confidence intervals (1,000 iterations). First, we tested a simple mediation model, in which PVS burden served as the mediator between habitual sleep duration and a range of behavioral outcomes. Next, we evaluated two sequential mediation models to explore alternative causal pathways involving both brain volume and PVS:

Path A: Sleep → Brain Volume → PVS → Behavioral Outcome
Path B: Sleep → PVS → Brain Volume → Behavioral Outcome

Each sequential model was tested separately using cortical gray matter, subcortical gray matter, and cerebral white matter as the brain volume mediator. The average causal mediation effect (ACME), average direct effect (ADE), and total effect were estimated for each model. All models included age and sex as covariates. Mediation effects were considered significant if the 95% confidence interval of the ACME did not include zero. To facilitate interpretation across outcomes, we extracted standardized estimates for specifical path. False discovery rate (FDR) correction was applied to p-values within each analytic family to account for multiple comparisons. All reported mediation p-values are FDR-adjusted. For ACME estimates, the sign of β reflects the combined direction of path a and path b; therefore, a negative ACME indicates that shorter sleep predicts worse outcomes through the mediator(s), whereas a positive ACME indicates the opposite direction.

**Sensitive analysis**

To further refine the probabilistic PVS template and enhance the reliability of PVS detection, we conducted a sensitivity analysis by applying two additional voxel-wise probability thresholds: 0.1% and 0.5%. These thresholds excluded voxels with low inter-individual overlap, further reducing the likelihood of including spurious or participant-specific artifacts.

Using these more conservative templates, we recalculated the PVS burden for each participant and repeated all primary analyses—including group comparisons, correlations with sleep variables, and mediation models—to assess the robustness of our findings. The results remained consistent across threshold levels, supporting the stability and validity of the associations identified in the main analyses.

**Statistics**

All statistical analyses were conducted in R version 4.2.0. Group differences in PVS burden, brain volumes, neurocognitive performance, and mental health symptoms were assessed using linear models controlling for age and sex. Effect sizes for group comparisons were reported as Cohen's d, with 95% confidence intervals estimated via bootstrap resampling. Associations between sleep measures (i.e., total sleep duration and parent-reported sleep disturbance), PVS burden, and behavioral outcomes were tested using multiple linear regression, also controlling for age and sex. Effect sizes for regression models were reported as Cohen's f², computed using the effect size package in R. All p-values were two-tailed, with significance set at $p < 0.05$. Data visualization was conducted using the ggplot2 package.

**Results**

**Characteristic of sleep group on glymphatic system**

The SI group (0.0096 ± 0.0042) exhibited a significantly greater PVS burden ($\beta = -5.742 \times 10^{-4}$, $t = -5.843$, $p = 5.36 \times 10^{-9}$, Cohen's $d = 0.147$ [95% CI: 0.097, 0.196], Figure 1 Panel A)) compared to the SS group (0.0089 ± 0.0039). Furthermore, continuous analyses revealed a significant negative association between PVS burden and total sleep duration ($\beta = -2.17 \times 10^{-4}$, $t = -5.548$, $p = 3 \times 10^{-8}$, Cohen's $f^2 = 0.0027$, Figure 1 Panel B), as well as a significant positive association with subjective sleep disturbance ($\beta = 2.44 \times 10^{-5}$, $t = 4.111$, $p = 3.99 \times 10^{-5}$, Cohen's $f^2 = 0.0025$, Figure 1 Panel C). These findings indicate that both insufficient sleep duration and poorer sleep quality are associated with increased PVS burden.

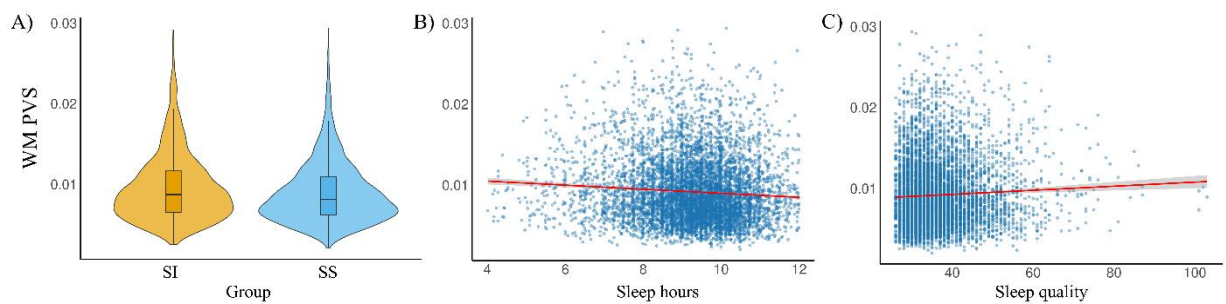

**Figure 1. Association between sleep and perivascular spaces (PVS) in white matter (WM).** A) Violin plot showing greater WM-PVS burden in the SI group compared to the SS group. B) Scatter plot illustrating a significant negative association between sleep duration and WM-PVS burden. C) Scatter plot showing a positive association between parent-reported sleep disturbance score and WM-PVS burden.

**Characteristic of sleep group on brain, cognition, mental health**

Adolescents in the SI group exhibited significantly lower brain volumes across multiple domains compared to their SS peers (Table S1). Specifically, significant reductions were observed in whole-brain cortical volume ($\beta = -9444$, $t = -7.27$, $p = 3.98 \times 10^{-13}$, Cohen's $d = 0.191$ [95% CI: 0.139, 0.243], Figure 2, Panel A), white matter volume ($\beta = -5444$, $t = -4.66$, $p = 3.23 \times 10^{-6}$, Cohen's $d = 0.122$ [95% CI: 0.070, 0.174], Panel B), and subcortical volume ($\beta = -669$, $t = -5.68$, $p = 1.38 \times 10^{-8}$,

Cohen's $d$ = 0.149 [95% CI: 0.097, 0.200], Panel C). Additionally, significant positive associations were observed between brain volume and normalized PVS burden. WM-PVS burden was positively correlated with cortical GM volume ($t$ = 3.37, $p$ = 0.0008, Cohen's $f^2$ = 0.0021, Figure 2, Panel D), white matter volume ($t$ = 6.64, $p$ = 3.31 × $10^{-11}$, $f^2$ = 0.0026, Panel E), and subcortical GM volume ($t$ = 3.85, $p$ = 0.0001, $f^2$ = 0.0021, Panel F).

Adolescents in the SI group demonstrated significantly lower neurocognitive performance across multiple domains compared to their SS peers (Figure 3 Panel A; Table S1). The most pronounced difference was observed in crystallized intelligence (β = 3.53 [2.60, 4.46], $p$ = 2.11 × $10^{-13}$, Cohen's $d$ = 0.20). The SS group also showed better performance in episodic memory, including immediate recall (β = 1.55 [0.91, 2.20], $p$ = 3.65 × $10^{-6}$, Cohen's $d$ = 0.11) and delayed recall (β = 0.28 [0.13, 0.42], $p$ = 2.9 × $10^{-4}$, Cohen's $d$ = 0.09), as well as in social cognition (β = 0.26 [0.18, 0.34], $p$ = 7.75 × $10^{-11}$, Cohen's $d$ = 0.16), visual processing speed (β = 0.17 [0.10, 0.25], $p$ = 1.52 × $10^{-5}$, Cohen's $d$ = 0.11), and executive function as measured by the Go/No-Go task (β = 0.81 [0.31, 1.31], $p$ = 0.0014, Cohen's $d$ = 0.08).

Furthermore, adolescents in the SI group exhibited significantly higher levels of psychopathological symptoms compared to their SS peers (Figure 3 Panel B; Table S1). The largest effect was observed in general psychopathology, as measured by the Brief Problem Monitor–Youth (BPM-Y: β = –0.10, 95% CI [–0.12, –0.09], $p$ < 1 × $10^{-16}$, Cohen's $d$ = –0.34) and the ABCL Total Problems scale (β = –1.35, 95% CI [–1.91, –0.79], $p$ = 2.21 × $10^{-6}$, $d$ = –0.13). The SI group also reported greater mood/bipolar symptoms (PGBI: β = –0.35, 95% CI [–0.48, –0.23], $p$ = 4.65 × $10^{-8}$, d = –0.14), increased peer relationship problems (PRP: β = –0.18, 95% CI [–0.22, –0.13], $p$ = 2.20 × $10^{-14}$, $d$ = –0.19), higher psychosis severity (PSS: β = 1.08, 95% CI [0.88, 1.29], $p$ < 1 × $10^{-16}$, $d$ = 0.25), and more psychotic-like experiences (PLEs: β = –0.21, 95% CI [–0.34, –0.08], $p$ = 0.0018, $d$ = –0.08). Additionally, the SI group showed elevated impulsivity-related traits, particularly negative urgency (UPPS: β = –0.56, 95% CI [–0.67, –0.45], $p < 1 \times 10^{-16}$, $d$ = –0.24).

In addition to the group differences between SI and SS, associations between sleep (both hours and quality) and behavioral outcomes (neurocognition, mental health) are presented in Figure S2 and S3.

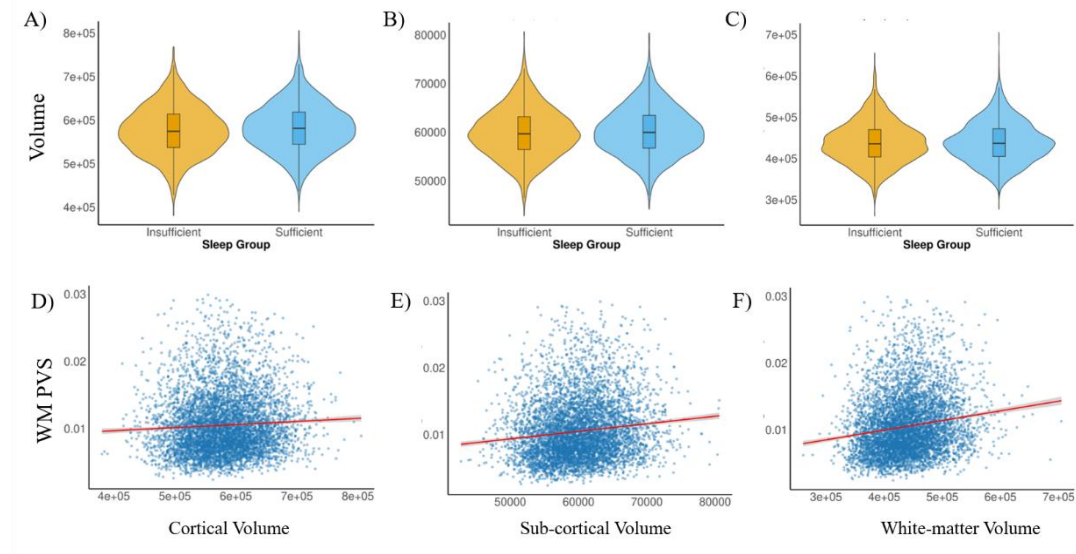

**Figure 2. Relationships between sleep sufficiency, brain volume, and PVS burden.** A–C) Violin plots depict group differences between sleep-sufficient and sleep-insufficient participants in (A) cortical brain volume, (B) subcortical gray matter volume, and (C) total cerebral white matter volume. Across all three metrics, the sleep-sufficient group shows higher brain volumes, suggesting a potential neuroanatomical pathway linking sleep quality to brain health. D–F) Scatterplots display linear relations (adjusted for age and sex) between WM PVS burden and (D) cortical gray matter volume, (E) subcortical gray matter volume, and (F) cerebral white matter volume. All three show small but significant positive associations, indicating that higher brain volumes are modestly associated with increased WM PVS burden in youth.

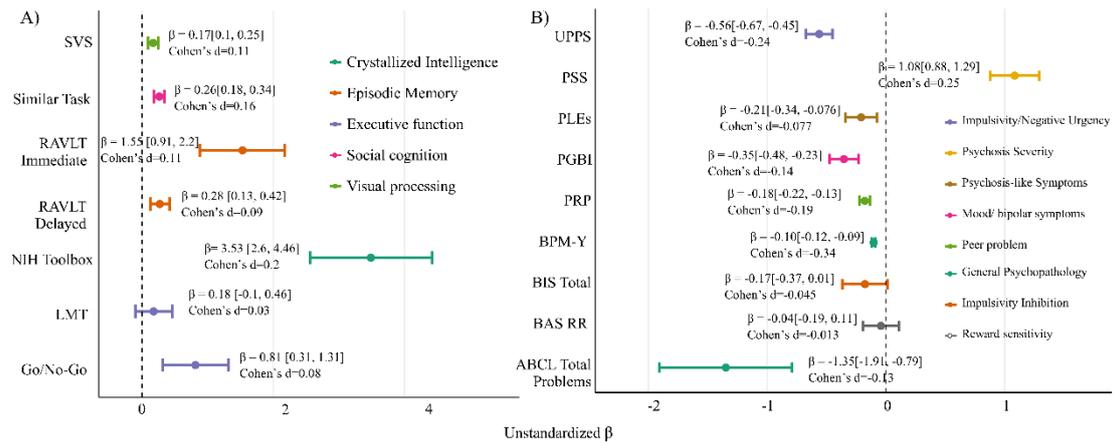

**Figure 3. Compared to SI, SS exhibited group differences in neurocognition and mental health outcomes.** A) Group differences in neurocognitive performance across five domains: crystallized intelligence, episodic memory, executive function, social cognition, and visual processing. B) Group differences in psychopathological symptoms across eight mental health dimensions. Each point represents the estimated unstandardized β coefficient from linear models comparing SS vs. SI, adjusted for age and sex, with 95% confidence intervals. Cohen's d effect sizes are displayed to the right of each point estimate. ABCL Total Problems: Achenbach Adult Behavior Checklist Total Problems Score; BAS RR: Behavioral Activation System – Reward Responsiveness; BIS Total: Behavioral Inhibition System Score; BMPY: General Psychopathology (BPM-Y); Go/No-Go: Go/No-Go Response Inhibition Task; LMT: List Memory Task; NIH Toolbox: NIH Toolbox Crystallized Composite Score; PGBI: Parenting reported Bipolar Symptoms; PLEs: Psychotic-Like Experiences; PRP: Peer Relationship Problems; PSS: Psychosis Symptom Severity; RAVLT Delayed: Rey Auditory Verbal Learning Test – Delayed Recall; RAVLT Immediate: Rey Auditory Verbal Learning Test – Immediate Recall; SVS: Symbol Visualization Speed; UPPS: UPPS-P Negative Urgency (Impulsivity).

**Mediation Analysis**

Since sleep is biologically linked to the glymphatic system and plays a critical role in brain health, we conducted mediation analyses to examine whether PVS serve as a mediating pathway linking sleep and behavioral outcomes. As shown in Figure 4

Panel A and Table S2, significant indirect effects via PVS were observed for multiple cognitive and mental health domains. In particular, crystallized intelligence (β = 0.062, *p* = 0.004, 4.31% indirect proportion), episodic memory (RAVLT Immediate: β = 0.0344, *p* = 0.008, 6.65% indirect proportion; RAVLT Delayed: β = 0.011, *p* = 0.002, 10.9% mediation), executive function (Go/No-Go: β = 0.02, *p* = 0.049, indirect proportion of 5.8%), and visual processing (SVS: β = 0.005, *p* = 0.002, indirect proportion of 6.2%) exhibited robust mediation effects, suggesting that part of the sleep–cognition relationship may operate through impaired glymphatic clearance. In the mental health domain, significant indirect effects were observed for psychosis severity (PSS: β = 0.013, *p* = 0.002; 2.51% mediated). These findings highlight that part of the sleep–cognition and sleep-mental health relationship may be mediated through impaired glymphatic clearance. Path models (Figure 4, Panel B) showed consistent a-path effects from sleep to PVS burden and b-path effects from PVS burden to cognitive and psychiatric outcomes. Additionally, we explored the mediation effect of sleep quality on behavioral alterations (Figure S4).

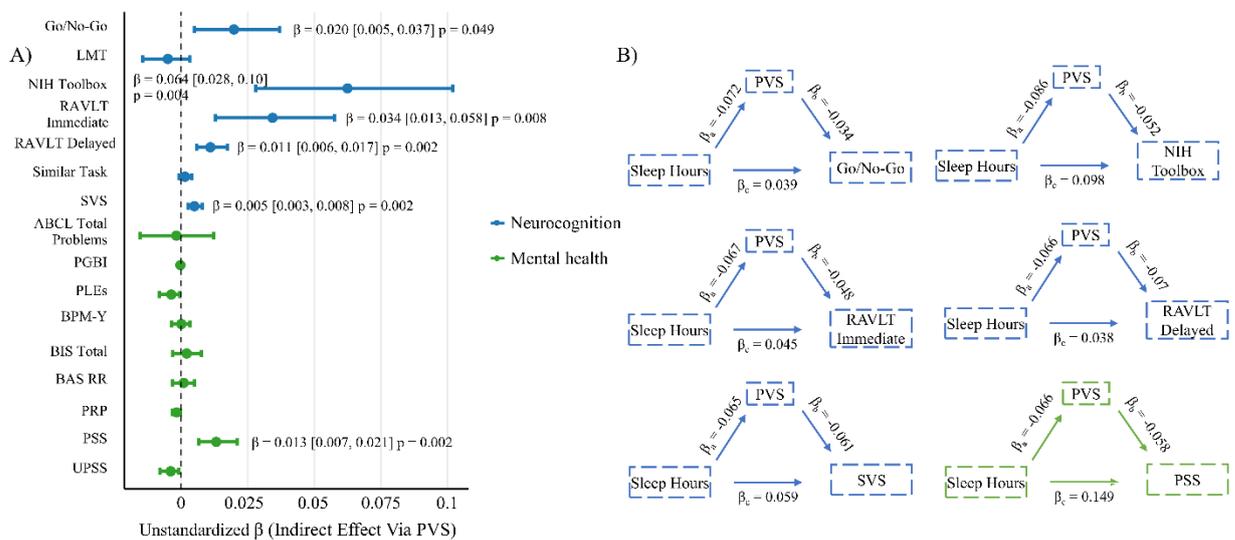

**Figure 4. Mediation effects of PVS on the relationship between sleep duration and behavioral outcomes.** A) Forest plot showing significant indirect effects (ACME) of sleep hours on cognitive and mental health outcomes via PVS burden. Outcomes are grouped by domain (blue: neurocognition; green: mental health), with horizontal bars representing 95% confidence intervals. B) Path diagrams for significant outcomes illustrate the mediation framework: sleep hours negatively predicted PVS burden (path a), which in turn predicted poorer behavioral outcomes (path b). Indirect effects (a × b) were significant in multiple domains, supporting the role of glymphatic dysfunction as a mechanistic pathway linking insufficient sleep with brain and behavioral changes during adolescence. All beta values are normalized. All the significant statistic after FDR corrections are noted in F

For sequential mediation with both PVS and brain volume as mediators, no significant effects were observed after p correction when cortical volume was used (Figure 5, Panel A).

For Path A, when using white matter volume as the initial mediator (Figure 5, Panel B) , significant indirect effects (ACME) were observed for several neurocognitive outcomes, including: executive function ($\beta$ = –0.00198, $p$ = 0.046), crystallized intelligence ($\beta$ = –0.0069, $p$ = 0.0038), episodic memory (Immediate: $\beta$ = –0.00364, p = 0.0042; Delayed: $\beta$ = –0.00116, $p$ = 0.0038), and visual processing ($\beta$ = –0.0053, $p$ = 0.0038). For mental health outcomes, significant effects were observed for: psychosis severity ($\beta$ = -0.0016, $p$ = 0.0038), impulsivity ($\beta$ = 0.0005, $p$ = 0.019). Subcortical volume yielded a similar pattern of effects in crystallized intelligence, episodic memory, visual processing, and psychosis severity (Figure 5, Panel C).

For Path B, where PVS burden precedes brain volume, more robust and widespread mediation effects were found (Figure 5, Panels B and C). Significant indirect effects included: executive function ($\beta$ = –0.003, $p$ = 0.008), crystallized intelligence ($\beta$ = –0.028, $p$ = 0.00017), episodic memory (Immediate: $\beta$ = –0.0024, $p$ = 0.033; Delayed: $\beta$ = –0.0008, $p$ = 0.01), social cognition ($\beta$ = –0.0002, $p$ = 0.044), and visual processing ($\beta$ = –0.0008, $p$ = 0.0022). For mental health, significant indirect effects in Path B were observed for: general psychopathology ($\beta$ = 0.00007, $p$ = 0.012), bipolar symptoms ($\beta$ = 0.0011, $p$ = 0.0022), psychosis severity ($\beta$ = –0.0018, $p$ = 0.002), and impulsivity ($\beta$ = 0.0006, $p$ = 0.01). These findings remained consistent when using subcortical volumes as mediators (Figure 5, Panel C), supporting the hypothesis that the pathway from sleep to PVS, followed by brain development and behavior, may represent a stronger and more plausible route.

Path diagrams for all outcomes with significant ACME effects are provided in the Figure S5–S8, following the format shown in Figure 5, Panel D.

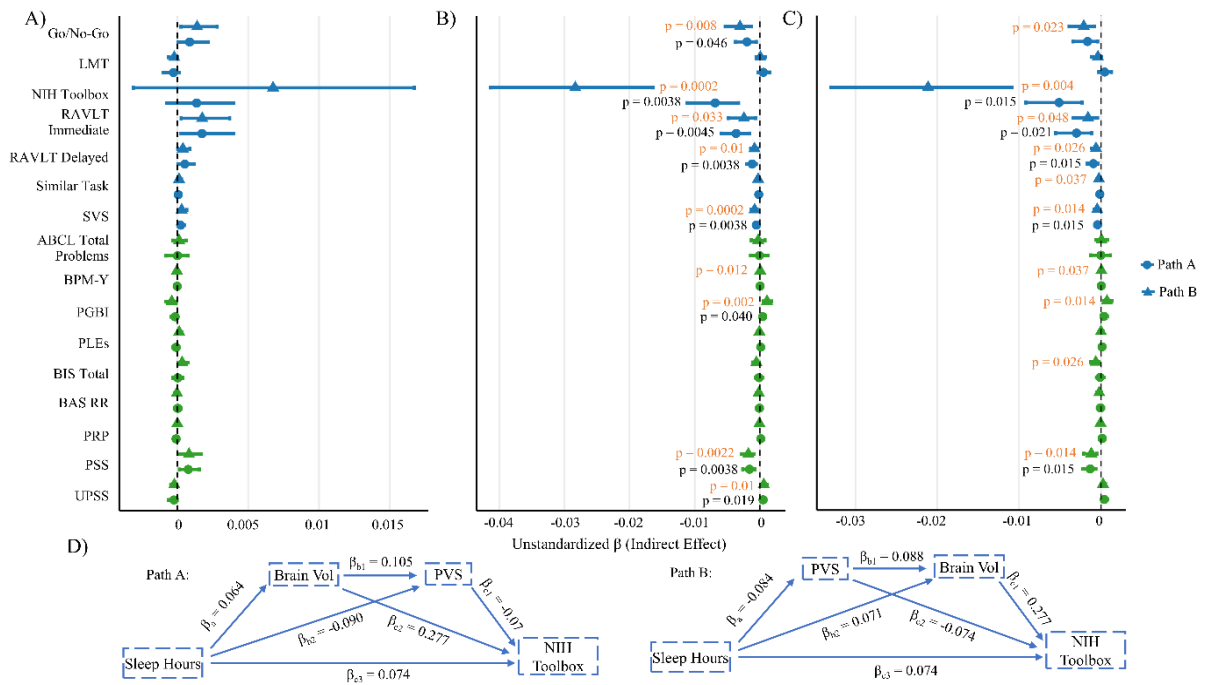

**Figure 5. Sequential mediation analysis results showing indirect effects (ACME β) of sleep on behavioral outcomes through brain structure and PVS.** A–C) Forest plots illustrate the indirect effects (ACME) of two sequential mediation pathways. Panel A shows results when using cortical volume as the brain measure, panel B for white matter volume, and panel C for subcortical volume. Each behavioral outcome is grouped by domain and plotted with 95% confidence intervals. D) Schematic example of sequential mediation pathways illustrating the influence of sleep hours on crystallized intelligence (NIH Toolbox), via either brain white matter volume and PVS (Path A) or PVS and brain white matter volume (Path B). Standardized β values and significance levels are shown for each pathway segment.

**Reproducible results with different threshold**

To evaluate the robustness of our findings, we conducted a series of sensitivity analyses using increasingly stringent definitions of PVS burden based on probabilistic template thresholds of 0.1% and 0.5% (Table S3).

At the 0.1% threshold, the PVS burden remained highly correlated with the original measure (r = 0.9995; Figure 6, Panel A). The SI group continued to exhibit significantly higher PVS burden ($\beta = -5.368 \times 10^{-4}$, t = –5.677, p = $1.42 \times 10^{-8}$, Cohen's d = 0.143 [95% CI: 0.09, 0.192]), and significant associations with sleep metrics were preserved (sleep duration: $\beta = -2.06 \times 10^{-4}$, t = –5.479, p = $4.42 \times 10^{-8}$, $f^2$ = 0.0025; sleep disturbance: $\beta = 2.252 \times 10^{-5}$, t = 3.937, p = $4.42 \times 10^{-5}$, $f^2$ = 0.0023).

At the 0.5% threshold, correlations with the original PVS measure remained strong (r = 0.983; Figure 6, Panel B). The SI group still demonstrated elevated PVS burden ($\beta = -4.363 \times 10^{-4}$, t = –5.447, p = $5.31 \times 10^{-8}$, Cohen's d = 0.137 [95% CI: 0.088, 0.186]), alongside robust associations with shorter sleep duration ($\beta = -1.698 \times 10^{-4}$, t = –5.33, p = $1.01 \times 10^{-7}$, $f^2$ = 0.0015) and greater sleep disturbance ($\beta = 2.11 \times 10^{-5}$, t = 4.321, p = $1.32 \times 10^{-5}$, $f^2$ = 0.0014).

Mediation models using PVS values derived from these conservative thresholds yielded consistent indirect effects in key domains (Figure S9-15), including crystallized intelligence, episodic memory, visual processing, and psychosis severity, reinforcing the stability of the findings. Collectively, these results demonstrate that the observed associations and mediation effects are robust across varying definitions of PVS burden.

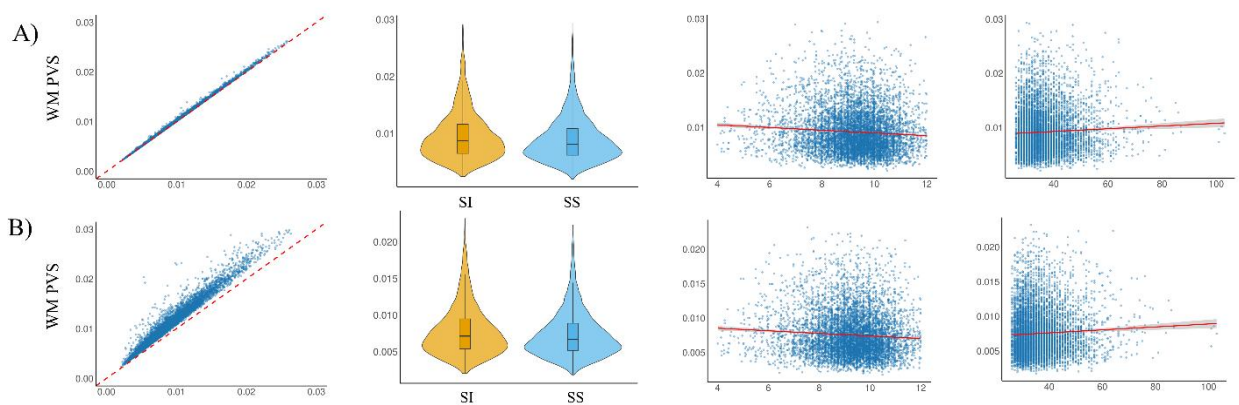

**Figure 6. Sensitivity analysis of WM PVS burden using alternative thresholds.** Panel A: probabilistic template thresholds 0.1%. Panel B: probabilistic template thresholds 0.5%. Each Panel showed comparisons between original and threshold WM PVS measures (first column), group differences between SI and SS (second column), and associations with sleep hours (third column) and sleep quality (fourth column).

## Discussion

The present study provides novel evidence that insufficient sleep during adolescence is associated with impaired glymphatic system (GS) function, as indexed by increased perivascular space (PVS) burden, and that this impairment partially mediates the adverse effects of poor sleep on brain structure, cognition, and mental health. These findings advance our understanding of the neurobiological mechanisms underlying sleep-related vulnerabilities during adolescence, and support the central hypothesis that the glymphatic pathway, essential for metabolic waste clearance in the central nervous system, serves as a crucial mediator of the detrimental effects of poor sleep during this sensitive developmental period.

**Shorter sleep duration and lower sleep quality during adolescence were associated with greater PVS burden.**

Although the relationship between sleep and GS function has been well characterized in animal models and adult populations, developmental trajectories and their behavioral implications remain poorly understood. Our finding that shorter sleep duration and lower sleep quality are associated with greater PVS burden aligns with the broader literature suggesting that sleep plays a critical role in glymphatic clearance. Specifically, sleep has been shown to facilitate the removal of neurotoxic waste products and metabolic byproducts via the GS, particularly during slow-wave sleep (Czeisler & Klerman, 1999; Imeri & Opp, 2009; Mignot, 2008; Robles & Carroll, 2011; Xie et al., 2013) via the GS(Iliff et al., 2014; Iliff et al., 2013; Iliff & Nedergaard, 2013; Iliff et al., 2012; Kress et al., 2014; Nedergaard, 2013). Impairments in this system may lead to, or be reflected by, enlarged PVS—a structural marker increasingly recognized as indicative of glymphatic dysfunction. Given the heightened metabolic demands of the adolescent brain, impaired clearance may have amplified consequences during this stage, reinforcing the importance of adequate sleep for maintaining neurophysiological homeostasis.

Although evidence for the sleep versus PVS associations in adolescents is limited, our sleep–PVS findings are consistent with prior studies in adults. In adults with sleep or neurological disorders, multiple studies (Berezuk et al., 2015; Opel et al., 2019; Si et al., 2020; Song et al., 2017) have reported associations between poor sleep characteristics and elevated PVS burden. Among healthy adults (n=552), Baril et al. (Baril et al., 2022) reported that lighter sleep—characterized by increased N1 sleep (light sleep) and decreased N3 sleep (deep sleep)—was associated with greater white matter PVS burden. However, two studies have reported mixed findings. In middle-aged and elderly adults (n = 559), Lysen et al. (Lysen et al., 2022) found a positive correlation between PVS load in the centrum semiovale and sleep efficiency but no correlations between sleep duration and PVS load. Similarly, Ramirez et al. (Ramirez et al., 2021), studying patients with cerebrovascular disease (n = 152), reported that longer total sleep time was associated with greater PVS volume, although this appeared to be mediated by time spent in bed rather than sleep per se. These discrepancies may reflect methodological differences, heterogeneous populations, or the complexity of sleep architecture in relation to PVS burden. Nonetheless, they suggest that the relationship between sleep and PVS is multifaceted and potentially influenced by developmental stage, health status, and specific aspects of sleep quality. Our findings contribute novel evidence supporting this relationship in adolescents.

**Insufficient sleep was linked to reduced brain volumes, poorer neurocognitive performance, and higher levels of psychopathologies**.

These findings are consistent with our previous studies based on the ABCD data (Turan et al., 2025; Yang et al., 2021; Yang, Liu, et al., 2022; Yang et al., 2023; Yang, Xie, et al., 2022a, 2022b), and underscore the pervasive influence of sleep on neurodevelopment and mental health.

**Normalized PVS volume was positively correlated with brain volume**.

After controlling white matter volume, PVS burden was correlated with cortical volume, white matter volume, and the subcortical volume and the correlation with white matter volume was the highest. This positive correlation suggests a functional necessity: larger tissue volume generates a need for a proportionally larger or more efficient system to handle the increased load of metabolic byproducts. The PVS is the structural correlate of this required clearance capacity. While these associations could be either compensatory or pathological, warranting further investigation, they highlight the complex interplay between GS function and brain morphology.

**The impact of sleep on various cognitive and mental health outcomes was significantly mediated by PVS burden, suggesting that the glymphatic pathway may partially account for the effects of sleep on brain and behavioral health**. Sequential mediation models further suggested that the pathway from sleep → PVS → brain volume → behavior was more robust than the reverse, supporting the hypothesis that glymphatic impairment may precede structural and functional alterations. These findings provide mechanistic insight into how insufficient sleep exerts long-term effects on brain and behavioral health, though causal animal models will be necessary to prove this potential change sequence.

**Limitations and Future Directions**

Several limitations warrant consideration. First, PVS burden was used as an indirect proxy for GS function; while widely accepted, it does not capture dynamic clearance processes. Technically, PVS burden estimation is subject to the cutoff threshold. Our results showed that changing the threshold did not change the statistical analysis results, reflecting a relatively homogeneous effects of the threshold on PVS burden estimation for all individuals. Second, the cross-sectional design precludes causal inference; longitudinal analyses are needed to confirm developmental trajectories. Third, sleep measures relied on questionnaires rather than objective metrics such as polysomnography or actigraphy, which could provide more granular insights into sleep architecture. Future studies should integrate multimodal imaging (e.g., DTI-ALPS, CSF flow MRI) and physiological measures to comprehensively characterize GS function. Additionally, examining interactions with genetic, environmental, and lifestyle factors will be critical for understanding individual differences in vulnerability and resilience. More importantly, future research should explore whether interventions targeting sleep hygiene or GS enhancement (e.g., optimizing posture, hydration, or pharmacological modulation) can reverse or prevent these adverse outcomes.

In summary, Adolescence represents a window of heightened vulnerability and opportunity for intervention. Our findings suggest that promoting sufficient and high-quality sleep may help preserve glymphatic function, thereby mitigating risks for cognitive deficits and psychopathology. Moreover, PVS burden could serve as a potential imaging biomarker for early detection of glymphatic dysfunction and related neurodevelopmental risks.

# Acknowledgment

Research efforts (data analysis and interpretation) in this work were supported by NIH grants: R01AG070227, R01AG081693, 1UL1TR003098. The preprint of this manuscript is available at doi:xxx. We thank the ABCD consortium and NIH for providing the data for performing the research in this work. Data used in the preparation of this article were obtained from the ABCD Study (https://abcdstudy.org/) and are held in the NIMH Data Archive. This is a multisite, longitudinal study designed to recruit more than 10,000 children aged 9–10 and follow them over 10 years into early adulthood. The ABCD Study is supported by the National Institutes of Health (NIH) and additional federal partners under award numbers U01DA041022, U01DA041028, U01DA041048, U01DA041089, U01DA041106, U01DA041117, U01DA041120, U01DA041134, U01DA041148, U01DA041156, U01DA041174, U24DA041123, and U24DA041147. A full list of supporters is available at https://abcdstudy.org/federal-partners/. A listing of participating sites and a complete listing of the study investigators can be found at https://abcdstudy.org/principal-investigators/. ABCD consortium investigators designed and implemented the study and/or provided data but did not necessarily participate in the analysis or writing of this report. This manuscript reflects the views of the authors and may not reflect the opinions or views of the NIH or ABCD consortium investigators. The authors are not paid to write this article by a pharmaceutical company or other agency. ABCD study received ethical approval in accordance with the ethical standards of the 1964 Declaration of Helsinki.

Curtis, S. (1992). Promoting health through a developmental analysis of adolescent risk behavior. *Journal of School Health*, *62*(9), 417-420.

Czeisler, C. A., & Klerman, E. B. (1999). Circadian and sleep-dependent regulation of hormone release in humans. *Recent Prog Horm Res*, *54*, 97-130; discussion 130-132. https://www.ncbi.nlm.nih.gov/pubmed/10548874

DiClemente, R. J., Hansen, W. B., & Ponton, L. E. (2013). *Handbook of adolescent health risk behavior*. Springer Science & Business Media.

Diekelmann, S., & Born, J. (2010). The memory function of sleep. *Nature Reviews Neuroscience*, *11*(2), 114-126.

Dumontheil, I., Burgess, P. W., & Blakemore, S. J. (2008). Development of rostral prefrontal cortex and cognitive and behavioural disorders. *Developmental Medicine & Child Neurology*, *50*(3), 168-181.

Eide, P. K., Vinje, V., Pripp, A. H., Mardal, K.-A., & Ringstad, G. (2021). Sleep deprivation impairs molecular clearance from the human brain. *Brain : a journal of neurology*, *144*(3), 863-874.

Fultz, N. E., Bonmassar, G., Setsompop, K., Stickgold, R. A., Rosen, B. R., Polimeni, J. R., & Lewis, L. D. (2019). Coupled electrophysiological, hemodynamic, and cerebrospinal fluid oscillations in human sleep. *Science*, *366*(6465), 628-631. https://doi.org/10.1126/science.aax5440

Giedd, J. N. (2004). Structural magnetic resonance imaging of the adolescent brain. *Ann N Y Acad Sci*, *1021*, 77-85. https://doi.org/10.1196/annals.1308.009

Glasser, M. F., Sotiropoulos, S. N., Wilson, J. A., Coalson, T. S., Fischl, B., Andersson, J. L.,…Jenkinson, M. (2013). The minimal preprocessing pipelines for the Human Connectome Project. *Neuroimage*, *80*, 105-124. https://doi.org/10.1016/j.neuroimage.2013.04.127

Gregory, A. M., & Sadeh, A. (2016). Annual research review: sleep problems in childhood psychiatric disorders–a review of the latest science. *Journal of child psychology and psychiatry*, *57*(3), 296-317.

Harrison, I. F., Ismail, O., Machhada, A., Colgan, N., Ohene, Y., Nahavandi, P.,…Lythgoe, M. F. (2020). Impaired glymphatic function and clearance of tau in an Alzheimer's disease model. *Brain : a journal of neurology*, *143*(8), 2576-2593. https://doi.org/10.1093/brain/awaa179

Hauglund, N. L., Andersen, M., Tokarska, K., Radovanovic, T., Kjaerby, C., Sørensen, F. L.,…Kolmos, M. G. (2025). Norepinephrine-mediated slow vasomotion drives glymphatic clearance during sleep. *Cell*, *188*(3), 606-622.e617.

Hauglund, N. L., Pavan, C., & Nedergaard, M. (2020). Cleaning the sleeping brain–the potential restorative function of the glymphatic system. *Current Opinion in Physiology*, *15*, 1-6.

Hobson, J. A. (2005). Sleep is of the brain, by the brain and for the brain. *Nature*, *437*(7063), 1254-1256.

Hussain, R., Tithof, J., Wang, W., Cheetham-West, A., Song, W., Peng, W.,…Peng, S. (2023). Potentiating glymphatic drainage minimizes post-traumatic cerebral oedema. *Nature*, *623*(7989), 992-1000.
27

Zhuo, J., Raghavan, P., Li, J., Roys, S., Njonkou Tchoquessi, R. L., Chen, H.,…Badjatia, N. (2024). Longitudinal assessment of glymphatic changes following mild traumatic brain injury: Insights from perivascular space burden and DTI-ALPS imaging [Original Research]. *Volume 15 - 2024*. https://doi.org/10.3389/fneur.2024.1443496

Zhuo, J., Raghavan, P., Shao, M., Roys, S., Liang, X., Tchoquessi, R. L. N.,…Gullapalli, R. P. (2024). Automatic Quantification of Enlarged Perivascular Space in Patients With Traumatic Brain Injury Using Super-Resolution of T2-Weighted Images. *J Neurotrauma*, *41*(3-4), 407-419. https://doi.org/10.1089/neu.2023.0082
33